\documentclass[12pt]{article}
\usepackage{amsmath}
\usepackage{graphicx,psfrag,epsf}
\usepackage{enumerate}
\usepackage{natbib}
\usepackage{url} 
\usepackage{float}
\usepackage[colorlinks=true,
            linkcolor=black,
            citecolor=black,
            filecolor=black,
            menucolor=black,
            runcolor=black,
            urlcolor=blue]{hyperref}

\newcommand{\blind}{0}

\newcommand{\beginsupplement}{%
        \setcounter{table}{0}
        \renewcommand{\thetable}{S\arabic{table}}%
        \setcounter{figure}{0}
        \renewcommand{\thefigure}{S\arabic{figure}}%
     }

\begin{document}

\def\spacingset#1{\renewcommand{\baselinestretch}%
{#1}\small\normalsize} \spacingset{1}


\if0\blind
{
 \title{\bf Sequential monitoring using the Second Generation P-Value with Type I error controlled by monitoring frequency}
 \author{Jonathan J. Chipman\thanks{
  This work was supported the support and resources from the Center for High Performance Computing at the University of Utah, the Advanced Computing Center for Research and Education at Vanderbilt University, and the grant NIDDK R01 DK100694.}\hspace{.2cm}\\
  Division of Biostatisitics,\\University of Utah Intermountain Healthcare\\Department of Population Health Sciences, \\ University of Utah\\
  Cancer Biostatistics, Huntsman Cancer Institute, University of Utah \\
  Robert A. Greevy, Jr. \\
  Department of Biostatistics, Vanderbilt University Medical Center, Vanderbilt \\
  Lindsay Mayberry \\
  Department of Medicine, Vanderbilt University Medical Center, Vanderbilt \\
  Jeffrey D. Blume\\
  School of Data Science, University of Virginia}
 \maketitle
} \fi

\if1\blind
{
 \bigskip
 \bigskip
 \bigskip
 \begin{center}
  {\LARGE\bf Title}
\end{center}
 \medskip
} \fi

\bigskip
\begin{abstract}
Many adaptive monitoring schemes adjust the required evidence toward a hypothesis to control Type I error. This shifts focus away from determining scientific relevance with an uncompromised degree of evidence. We propose sequentially monitoring the Second Generation P-Value (SGPV) on repeated intervals until establishing evidence for scientific relevance (SeqSGPV). SeqSGPV encompasses existing strategies to monitor Region of Practical Equivalence (ROPE) or Region of Equivalence (ROE) hypotheses. Hence, our focus is to formalize sequential SGPV monitoring; establish a novel set of scientific hypotheses, called PRISM, which is a ROE with a ROPE surrounding the null hypothesis; and use monitoring frequency and a novel affirmation step to control Type I error. Under immediate and delayed outcomes, we assess finite and limiting SeqSGPV operating characteristics when monitoring PRISM, ROPE, and null-bound ROE hypotheses. In extensive simulations, SeqSGPV PRISM monitoring reduced wait time for fully sequential monitoring, average sample size, and reversals of null hypothesis conclusions under the null. With real-world data, we design a SeqSGPV-monitored randomized trial. SeqSGPV is method-agnostic and easy to implement. Adjusting monitoring frequency/affirmation and monitoring a one-sided PRISM synergistically control Type I error. PRISM monitoring and adjusting monitoring frequency to control Type I error may have application beyond SeqSGPV.
\end{abstract}

\noindent%
{\it Keywords:} Second Generation P-Value, repeated interval monitoring, evidence monitoring, Pre-Specified Regions Indicating Scientific Merit, monitoring frequency, delayed outcomes
\vfill

\newpage
\spacingset{1.45} 
\section{Introduction}
\label{sec:intro}

The Second Generation P-Value (SGPV) \citep{Blume:SGPV,blume2019introduction} is an evidence-based metric that measures the overlap between an inferential interval and a composite hypothesis. By pre-specifying a set of effects deemed indifferent from the null hypothesis (i.e. indifference zone), the SGPV brings study design transparency and bridges the gap between statistical significance and scientific relevance. The SGPV has been incorporated into statistical methods and analyses to address common research objectives \citep{stewart2019second,Zuo2021,Sirohi2022}. Our objective is to develop a sequential monitoring scheme using the SGPV (SeqSGPV) for establishing evidence of an effect being scientifically relevant or irrelevant. The proposed design may be used with a maximum or unrestricted sample size while controlling the Type I error through monitoring frequency and/or a novel affirmation rule. Fitting within our framework are existing strategies to monitor scientifically relevant regions/hypotheses using repeated credible intervals \citep{Hobbs:2008ce,Kruschke:2013jy} or confidence intervals \citep{jennison1989interim}. Hence, our focus will be on formulation and general-purpose nature of SGPV monitoring, on a novel set of regions/hypotheses of meaningful effects, and on monitoring frequency mechanisms to control Type I error. The later two advancements may have broader application beyond SeqSGPV.

Calculating the SGPV requires a pre-specified composite hypothesis and an inferential interval. A confidence, credible, support, or other interval may be used, which motivates describing the SGPV as "method-agnostic" \citep{stewart2019second}. Rejecting a scientifically relevant indifference zone is stronger than statistical significance and provides protection against Type I error. It's Type I error protection contrasts common methods to control the family-wise error rate which rank-order p-values and use an $\alpha$-adjusted threshold to determine statistical significance. With p-value testing, trivial effects may be statistically significant whereas SGPV testing also requires the effect to be scientifically meaningful. For example, given the same family-wise error rate, the set of meaningful multiple comparisons meeting SGPV criteria are not necessarily the same as the set of comparisons meeting Bonferroni adjustments -- see \citet{Blume:SGPV} for a genomics example.

The indifference zone has also been termed a Region of Practical Equivalence (ROPE) \citep{Kruschke:2013jy} and a Region of Equivalence (ROE) \citep{Freedman:1984wz}. Whereas ROPE encompasses the null hypothesis, the more flexible ROE may encompass, exclude, or include the null hypothesis as a region boundary. The landmark umbrella trial BATTLE -- Biomarker-integrated approaches of targeted therapy of lung cancer elimination -- established treatment futility and efficacy as having, respectively, low probability of being scientifically meaningful and high probability of being better than the null \citep{Zhou2008}. Though not stated explicitly, this reflects a null-bound ROE. A design can indirectly induce a ROE by drawing inference on an effect being greater or less than a given effect such as the null hypothesis \citep{viele2020comparison} or the minimum scientifically meaningful effect \citep{Wathen2017}. The efficacy/futility probability criteria for each of the cited designs was tuned to achieve a desired false discovery rate which is a common strategy in Bayesian sequentially monitored designs \citep{thall1994practical, berry2006}. In an analogous fashion, Frequentist designs commonly adjust $\alpha$ across sequential monitoring evaluations \citep{lan1993sequential}. Hence, $\alpha$-adjusted Frequentist designs are induced null-bound ROEs that change as $\alpha$ changes.

In contrast, the Sequential Probability Ratio Test \citep{Wald:1945kd}, Sequential Bayes Factor \citep{Schonbrodt2018}, and monitoring ROPE/ROE with repeated credible intervals do not tune the inference-specific evidence required to favor one fixed hypothesis over another. Asymptotically, each of these designs converge to support the hypothesis that best supports the truth, though if two hypotheses equally support the true effect, neither hypothesis will be supported more than the other \citep{Blume2002}. The expected sample size to reach a degree of evidence asymptotes to infinity as the effects are more equally supported by each hypothesis. Truncating the sample size and the Triangular Test \citep{anderson1960modification} have been proposed to address this limitation. Although evidence-based designs may achieve a fixed level of evidence, monitoring frequency can differentially impact the Type I error. In any monitoring scheme, each additional interim assessment increases the risk of false discoveries; the added risk converges to zero when comparing two fixed hypotheses \citep{Blume:2008cw}. This makes adjusting monitoring frequency a targetable mechanism for controlling false discoveries without sacrificing evidence toward a hypothesis. 

Limited research focuses on monitoring timing and frequency to control Type I error and minimize average sample size.  Studies have focused on when to perform a single, or a few, interim analyses \citep{Togo2013,Wu2020} and on the impact of monitoring for futility, efficacy, or both upon Type I error and power \citep{Ryan2020}. To the best of our knowledge, there is not literature on setting the minimum wait time to ensure a nominal Type I error with fully sequential monitoring nor an investigation on adjusting monitoring frequency to reduce Type I error. Through simulation, we will demonstrate an interplay and potential synergy in controlling Type I error between monitoring frequency adjustments and indifference zone specification. This can lead to reduced average sample size and, in the presence of delayed outcomes, can reduce the risk of reversing conclusions under the null hypothesis.

Our development of SeqSGPV begins in Section~\ref{sec:prelim} by establishing a preliminary framework for evidence and scientifically meaningful effects. Along with notation, we re-introduce the SGPV and introduce a new set of scientifically meaningful regions/hypotheses called the Region of Meaningful Effects (ROME) and Pre-Specified Regions Indicating Scientific Merit (PRISM). We list possible PRISM conclusions based upon the SGPV. In Section~\ref{sec:prism}, we provide SeqSGPV rules for monitoring PRISM hypotheses. Emphasis is placed on controlling Type I error through monitoring frequency including a novel affirmation step. Simulations in Section~\ref{sec:sims} investigate limiting and finite sample error probabilities and the impact of monitoring frequencies in the presence of immediate and delayed outcomes. We compare SeqSGPV monitoring of PRISM hypotheses to monitoring ROPE and null-bound ROE hypotheses. Section~\ref{sec:reach} is a real-world application of SeqSGPV monitoring of PRISM hypotheses using data from a randomized trial on diabetes management. We end in Section~\ref{sec:conc} with conclusions and practical considerations.

\section{Preliminary framework and notation}
\label{sec:prelim}

We review the SGPV, introduce one- and two-sided PRISM hypotheses, and summarize possible SGPV PRISM conclusions.

\subsection{SGPV}

The SGPV indicates when the data are compatible with effects indifferent from the null or alternative hypothesis or when the data are inconclusive. More generally, it may be used for any composite hypothesis \textit{H}. The SGPV calculates the overlap between an interval \textit{I} and the set of effects $\mathrm{\Delta}$ in \textit{H}. The interval includes [\textit{a}, \textit{b}] where \textit{a} and \textit{b} are real numbers such that \textit{a} $<$ \textit{b}, and the length of the interval is \textit{b - a} and denoted $|I|$. The overlap between the interval and the set $\mathrm{\Delta}$ is $\left| I \cap H_{\Delta} \right|$. The SGPV is then calculated as

$$
p_{H} = \frac{| I \cap \Delta |}{|I|} \times \max \left\{\frac{| I |}{2 | \Delta |}, 1 \right\}.
$$

The adjustment, $max\left\{\frac{|I|}{2|H_{\Delta}|},1\right\}$, is a small sample size correction – setting $p_H$ to half of the overlap when the inferential interval overwhelms $H_{\Delta}$ by at least twice the length.

\subsection{PRISM}

\begin{figure}[H]
\centering
\includegraphics[width=1\textwidth]{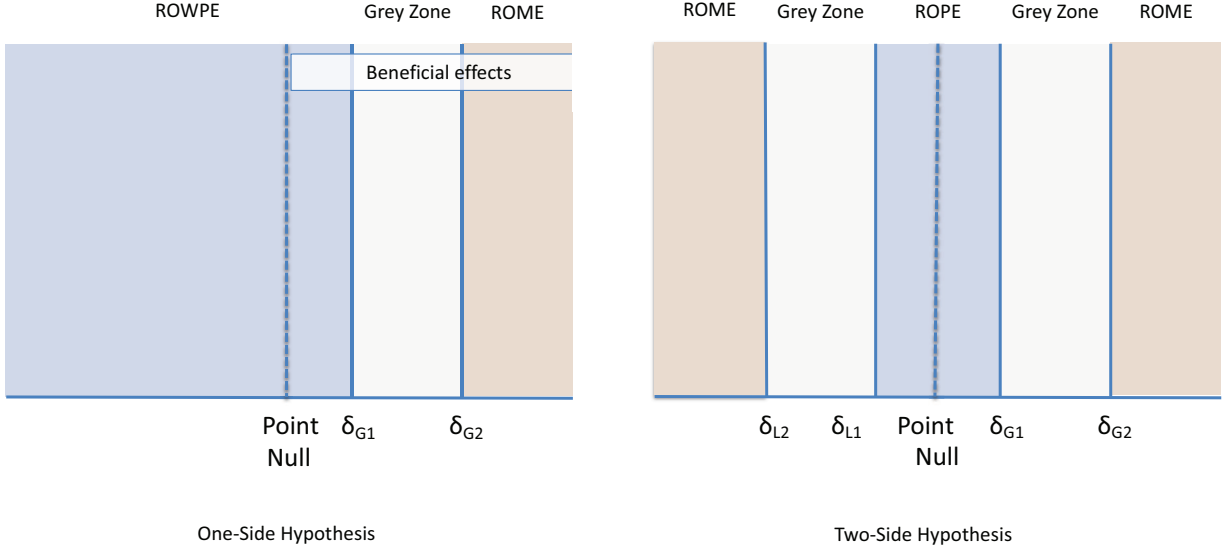}
\caption{\label{fig:clinicalGuideposts}Pre-Specified Regions Indicating Scientific Merit (PRISM) for one- and two-sided hypotheses. The PRISM always includes an indifference zone that surrounds the point null hypothesis (i.e. ROPE/ROWPE).}
\end{figure}

In a two-sided hypothesis, the PRISM includes a ROPE which surrounds the point null hypothesis and a ROME which is bounded by the minimum clinically meaningful effect. ROPE effects lay between $\delta_{L1}$ (essentially equivalent yet less than the point null) and $\delta_{G1}$ (essentially equivalent yet greater than the point null). Meaningful effects less than the point null are of a magnitude of at least of $\delta_{L2}$, while meaningful effects greater than the point null are of magnitude of at least $\delta_{G2}$. The remaining, mild effects reflect the ROE, which we also call the Grey Zone for exposition. A one-sided hypothesis also has a Grey Zone and ROME; however, instead of having a ROPE, it has a Region of Worse or Practically Equivalent (ROWPE) effects. When a positive effect is desirable, the ROWPE includes effects of magnitude worse than and up to $\delta_{G1}$ (Figure~\ref{fig:clinicalGuideposts}). The one-sided PRISM is more restrictive than a one-sided ROE by requiring that null effects not be on the ROE boundary.

We define and focus on a set of PRISM monitoring hypotheses.

\begin{itemize}
\item \textbf{ROPE / ROWPE Hypothesis:} The treatment effect lies within ROPE$_{[\delta_{L1},\delta_{G1}]}$ for a two-sided hypothesis, ROWPE$_{(-\infty,\delta_{G1}]}$ for a one-sided hypothesis where a positive effect is desirable, and ROWPE$_{[\delta_{L1},\infty)}$ where a negative effect is desirable.

\item \textbf{ROME Hypothesis:} The treatment effect lies within ROME$_{(-\infty,\delta_{L2}] \cup [\delta_{G2},\infty)}$ for a two-sided hypothesis, ROME$_{[\delta_{G2},\infty)}$ for a one-sided hypothesis where a positive effect is desirable, and ROME$_{(-\infty, \delta_{L2}]}$ where a negative effect is desirable.
\end{itemize}

\begin{figure}[H]
\centering
\includegraphics[width=1\textwidth]{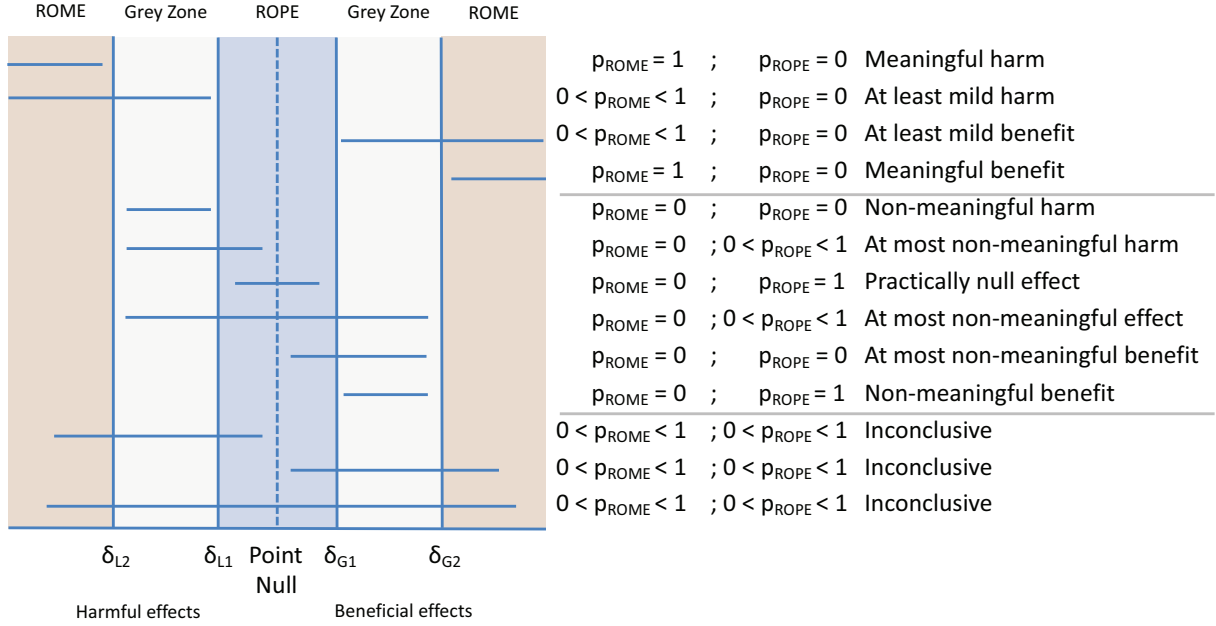}
\caption{\label{fig:prismConclusions}Possible two-sided PRISM SGPV conclusions. Similar conclusions may be drawn upon a one-sided PRISM. SeqSGPV monitors for conclusive evidence to rule out essentially null effects (p$_{ROPE}$=0 or p$_{ROWPE}$=0) or to rule out meaningful effects (p$_{ROME}$=0).}
\end{figure}

For a ROPE null hypothesis, practically equivalent effects are ruled out when p$_{ROPE}$=0 whereas the data support practically null effects when p$_{ROPE}$=1. The data are inconclusive when 0$<p_{ROPE}<$1. The interpretation is similar for a ROME hypothesis. Neither of these hypotheses include mild effects in the Grey Zone. When applied to a two-sided hypothesis, thirteen conclusions may be drawn from the SGPV (Figure~\ref{fig:prismConclusions}); eight conclusions may be drawn for a one-sided hypothesis. For sequential monitoring, we reduce to three conclusions:

\begin{itemize}
  \item When $p_{ROME}=0$, scientifically meaningful treatment effects are ruled out.
  \item When $p_{ROPE}=0$, effects essentially equivalent to the null are ruled out.
  \item Otherwise, the treatment effect is of inconclusive merit.
\end{itemize}

\section{SeqSGPV}
\label{sec:prism}

We introduce sequential monitoring rules for measuring SGPV evidence on PRISM hypotheses. Achieving a fixed level of evidence and controlling Type I errors are separate objectives. Evidence is measured through the SGPV, and a Type I error occurs when the final interval excludes the null hypothesis. Adjusting monitoring frequency is a means of obtaining a fixed level of evidence while controlling Type I error. 

\subsection{Monitoring frequency to control Type I error}

The suite of monitoring frequency parameters include an initial wait time before the first assessment ($W$), the number of steps or observations between assessments ($S$), a number of steps or observations required to affirm a stopping indication ($A$), and a maximum sample size ($N$) which may be set as unrestricted. The raised alert does not need to be consistent for $A$ consecutive subjects.

\subsection{SeqSGPV PRISM monitoring}

Monitoring PRISM hypotheses using the SGPV proceeds as follows.

\begin{enumerate}
  \item \textbf{Set PRISM:} Based on scientific relevance, a-priori determine the PRISM with ROME and ROPE for a two-sided hypothesis or ROME and ROWPE for a one-sided hypothesis.
  \item \textbf{Set Monitoring Frequency:} Choose $W$, $S$, $A$, and $N$ to achieve desired operating characteristics.
  \item \textbf{Wait to Monitor:} Enroll $W$ observations before applying monitoring rules.
  \item \textbf{Monitor Evidence:} At a monitoring frequency, $S$, calculate the SGPV using an inferential interval of choice. Raise an alert when $p_{ROME}=0$ or $p_{ROPE}=0$. Intervals used throughout monitoring are called monitoring intervals.
  \item \textbf{Affirm Alert:} Continue monitoring until affirming the same alert after an additional $A$ observations, i.e. $p_{ROME}$ or $p_{ROPE}$ still equals 0. Or, if using a backward-looking alert, stop if the same alert was raised the $A^{th}$ previous outcome.
  \item \textbf{Stop:} Stop at affirming an alert or at $N$.
\end{enumerate}

The design choice of $W$, $S$, $A$, and $N$ are determined through simulation under an assumed outcome variability. As data accumulate, $S$, $A$, and $N$ may be adapted to continually ensure Type I error control by reassessing the operating characteristics using the available data. Simulated examples are provided in Supplemental Figure~\ref{fig:supExamples}.

SeqSGPV captures the gradated merit of Grey Zone effects. Grey Zone effects are more likely to be found meaningful as they approach ROME and more likely to be found as essentially null as they approach ROPE. Mid-Grey Zone effects are equally likely to draw a conclusion as being non-ROPE or non-ROME and have the greatest chance of raising conflicting alerts – for example raising an alert for a non-ROPE effect then $A$ observations later an alert for a non-ROME effect. For this reason, the affirmation alert should be the same as the alert it affirms.

By the likelihood principle, credible and support intervals are not impacted by sampling intentions, whereas confidence intervals are constructed to maintain an overall Type I error across multiple evaluations. Because the proposed stopping rules are based upon SGPV evidence -- as compared to statistical significance -- a repeated 1-$\alpha$ confidence interval may be used as a monitoring interval. Only the final interval is to be interpreted and exact coverage is not expected across all treatment effects. Rephrased, the study design is developed for multiple evaluations of evidence and a single evaluation of statistical significance.

\section{Simulation studies}
\label{sec:sims}

We assess operating characteristics when monitoring PRISM versus ROPE-only monitoring (two-sided hypothesis) and versus null-bound ROE monitoring (one-sided hypothesis). Designs are comparable by achieving the same level of evidence towards hypotheses of interest while while controlling Type I error. Treatment effects are assume to be equally plausible though a distribution of true effects may be assumed \citep{Spiegelhalter:2004vw}.

\subsection{Limiting errors and risk of being inconclusive}
\label{subsec:limitingprobs}

To assess the limiting properties SeqSGPV, we replicated and expanded upon simulations performed by \citet{Kruschke:2013jy} which monitored a ROPE using repeated 95\% credible intervals. We compared the probability of rejecting the null hypothesis and of being SGPV inconclusive between monitoring ROPE versus PRISM hypotheses (two-sided hypothesis) and between null-bound ROE versus PRISM hypotheses (one-sided hypothesis).

We assumed a setting of immediate standard normal outcomes with treatment effects of 0, 0.15, 0.325, and 0.5. The two-sided PRISM included ROPE$_{[-0.15, 0.15]}$ and ROME$_{(-\infty, -0.5]\cup[0.5, \infty)}$ hypotheses and was compared to monitoring ROPE$_{[-0.15, 0.15]}$. The one-sided PRISM included ROWPE$_{(-\infty, 0.15]}$ and ROME$_{[0.5, \infty)}$ hypotheses and was compared to two null-bound ROE hypotheses: ROE$_{[0,0.15]}$ and ROE$_{[0,0.5]}$. ROPE monitoring stopped when ruling out ROPE ($p_{ROPE}=0$) or providing evidence for ROPE (i.e. p$_{ROPE}$=1). ROE monitoring stopped when ruling out effects greater than 0 (i.e. p$_{[0,\infty)}$=0) or when ruling out meaningful effects (i.e. p$_{(-\infty,\delta_1]}$=0). Hence, ROPE monitoring follows two non-overlapping hypotheses and ROE monitoring follows overlapping hypotheses. ROE hypotheses are structured similar as the BATTLE trial hypotheses.

Simulated observations were alternately randomized to treatment or standard of care. After each arm enrolled two observations, fully sequential monitoring began with calculating p$_{ROPE}$ and p$_{ROME}$ on a 95\% monitoring intervals with no required affirmation (monitoring frequencies: W=4, S=1, A=0, and N = $\infty$). Unadjusted 95\% confidence intervals were used as monitoring intervals, which is mathematically equivalent to a 6.83 support interval and essentially equivalent to a 95\% credible interval with a flat prior on the mean.

\begin{figure}[H]
\centering
\includegraphics[width=1\textwidth]{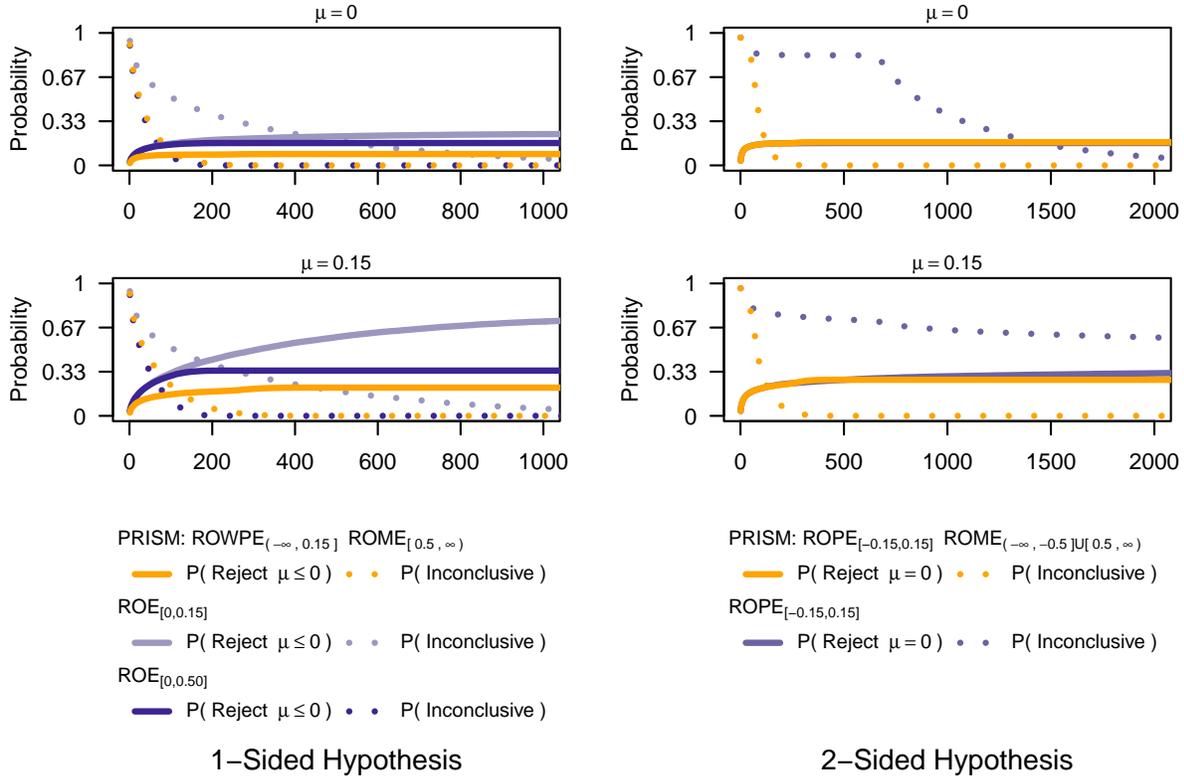}
\caption{\label{fig:limitingProbs}Limiting probability of rejecting the null hypothesis and of being inconclusive. Here, the null-bound ROEs are monitored until ruling out beneficial non-null effects (i.e. p$_{[0,\infty)}$=0) or non-meaningful effects (i.e. p$_{(-\infty, \delta_{G1}]}$=0). ROPE is monitored until ruling out ROPE (p$_{ROPE}$=0) or supporting ROPE (i.e. p$_{ROPE}$=1). See Figure~\ref{fig:prismConclusions} for PRISM monitoring rules and Supplemental Figure~\ref{fig:SuplimitingProbs} for limiting probabilities under other treatment effects.}
\end{figure}

\begin{figure}[H]
\centering
\includegraphics[width=1\textwidth]{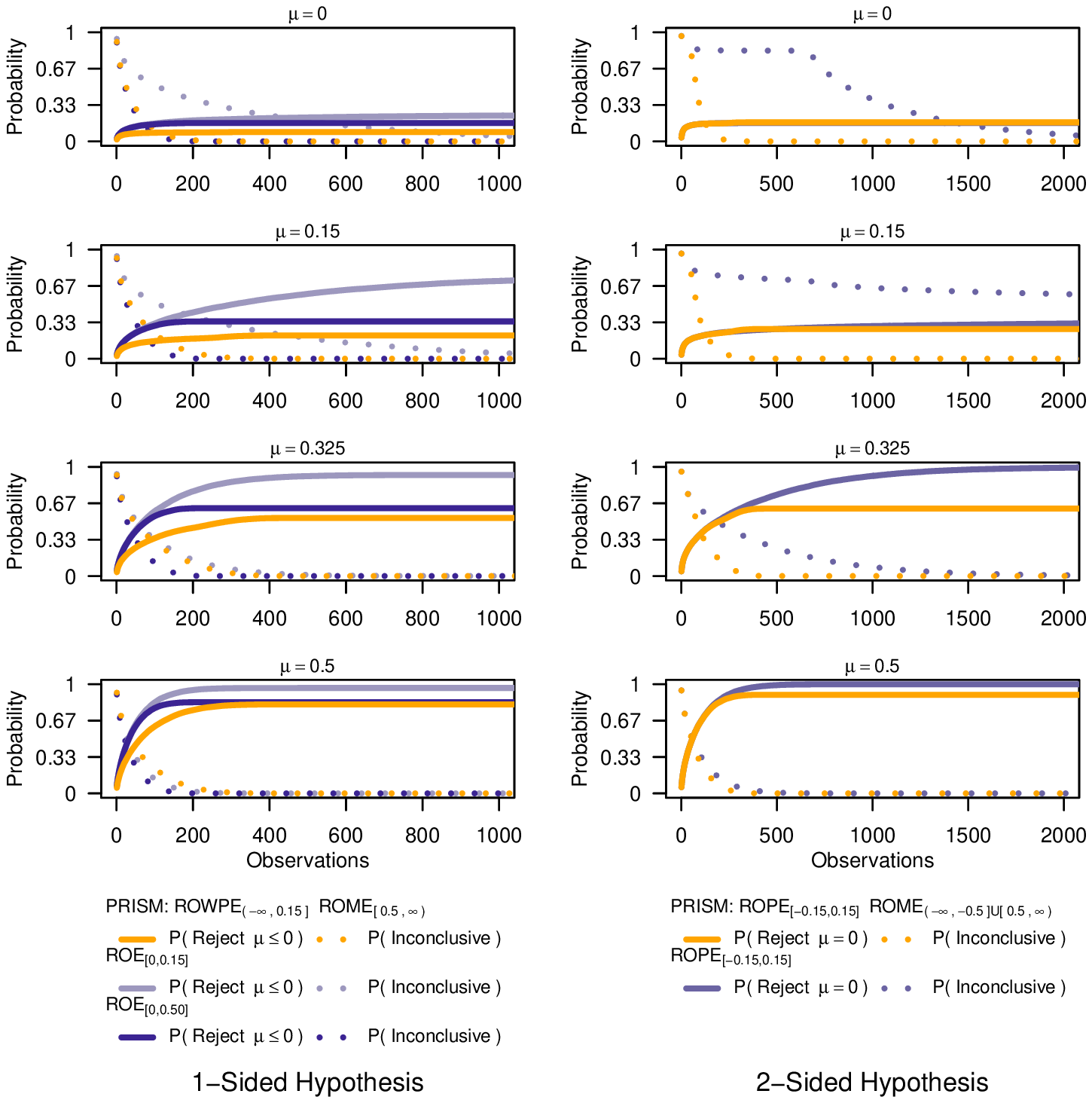}
\caption{\label{fig:SuplimitingProbs}Limiting probability of rejecting the null hypothesis and of being inconclusive. See Figure~\ref{fig:prismConclusions} for PRISM monitoring rules. Here, the null-bound ROEs are monitored until ruling out beneficial non-null effects (p$_{[0,\infty)}$=0) or non-meaningful effects (p$_{(-\infty, \delta_{G1}]}$=0). ROPE is monitored until ruling out ROPE (p$_{ROPE}$=0) or supporting ROPE (p$_{ROPE}$=1).}
\end{figure}

Under a one-sided hypothesis and treatment effect of 0, PRISM monitoring had a smaller limiting Type I error rate compared to monitoring either null-bound ROE hypotheses (Figure~\ref{fig:limitingProbs} and Supplemental Figure~\ref{fig:SuplimitingProbs}). At a treatment effect of 0, PRISM monitoring was almost as quickly conclusive as ROE$_{[0,0.5]}$ monitoring. PRISM monitoring required the most data to be conclusive in the middle of the Grey Zone, and it was less conclusive than monitoring the alternative ROE hypotheses when the treatment effect was 0.5. Acknowledging different Type I error rates, the change in inconclusivity was relatively greater in the PRISM design than alternatives.

Under a two-sided hypothesis, the limiting SeqSGPV Type I error was nearly equivalent between PRISM and ROPE-only monitoring. Yet, PRISM monitoring was dramatically more quickly conclusive than ROPE-only monitoring which, at the ROPE boundary, had a risk of never being conclusive. This was an artifact of the open sample size and of monitoring non-overlapping hypotheses. A continuous effect will never exactly correspond with a boundary; however, the expected sample size required to be conclusive approaches infinity as the effect approaches the boundary. PRISM's overlapping hypotheses (and the overlapping hypotheses set for monitoring ROE) resolved this limitation. The quickness of being conclusive is sensitive to the extent to which the monitored hypotheses overlap. Less overlap increases the time to being conclusive for effects within the overlapping hypothesis.

\subsection{Monitoring frequency to control Type I error}
\label{subsec:alphacontrol}
In practice, many studies will not have the luxury of unrestricted monitoring. A maximum sample size is typically specified, for which frequency properties are calculated. Here we look at the Type I error trajectory for different monitoring frequencies ($W$, $S$, and $A$) across maximum sample sizes when using the one- and two-sided PRISM, two-sided ROPE, and one-sided ROE hypotheses described in Section~\ref{subsec:limitingprobs}. 

400,000 Two-arm randomized trials were simulated, enrolling up to 5,000 observations equally randomized between the arms. Outcomes for both arms were standard normal distributed. For each of the PRISM, ROPE, and ROE boundaries, we considered monitoring frequencies of $S=1$ and $S=10$ and required affirmation steps of $A=0$ and $A=10$, and 100 different wait times based upon the amount of expected statistical information collected. The shortest wait time was 4 units for which the interval's expected half-width was 2 (i.e. 4 * absolute ROME boundary). And the longest wait time was 4700 units, roughly corresponding with an interval's expected half-width of 0.05 (i.e. 1/3 * absolute ROPE boundary). Unadjusted 95\% confidence intervals were used for monitoring intervals, and the maximum acceptable Type I error of interest was 0.05. Starting with the first look and for subsequent looks, we calculated the cumulative proportion of trials that had rejected the null hypothesis including whether the current look (the considered maximum sample size) would reject the null hypothesis. We estimated the average sample size under the earliest wait time while controlling Type I error.

\begin{figure}[H]
\centering
\includegraphics[width=1\textwidth]{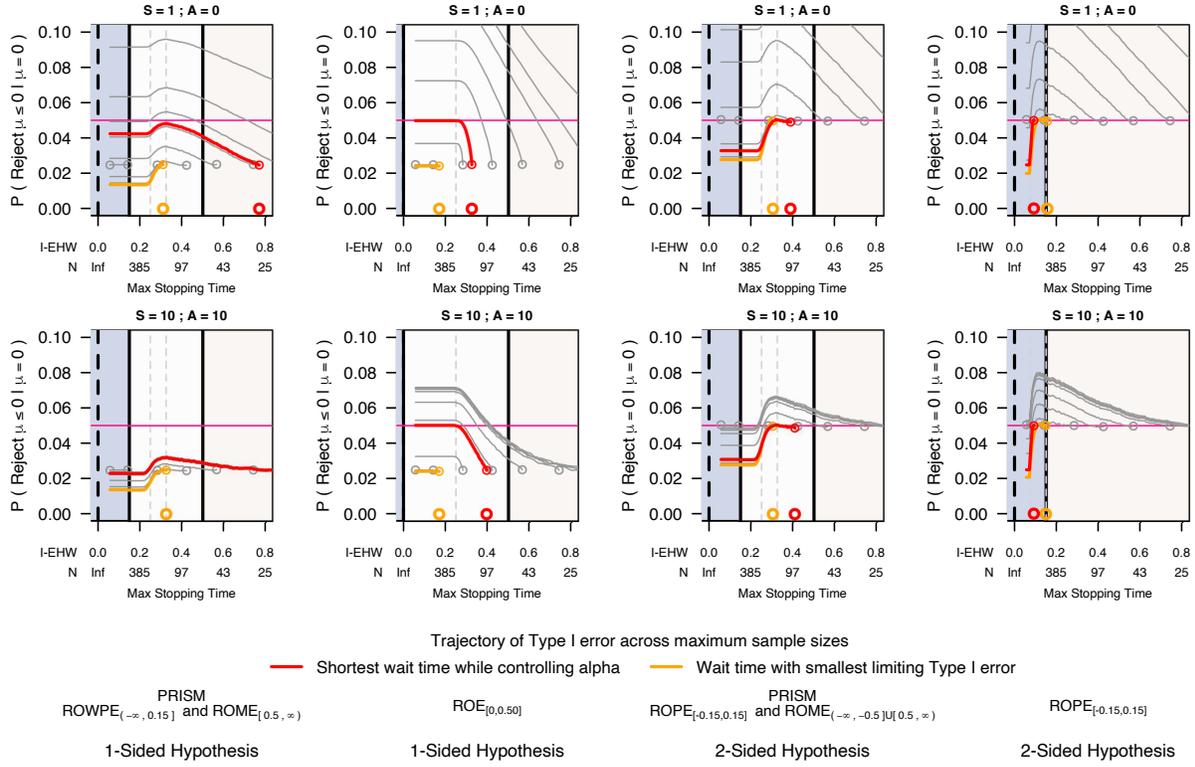}
\caption{\label{fig:t1etrajectory}Type I error as a function of wait time ($W$), monitoring steps ($S$), affirmation steps ($A$), and maximum sample size ($N$). The x axis reads from right to left in terms of accruing data and left to right in terms of interval expected half-width (I-EHW). Circles denote $W$ and Type I error generally increases for fully sequential monitoring until $N$. Dashed lines reflect the midpoint between 0 and the positive ROME boundary and between ROPE and ROME positive boundaries.}
\end{figure}

\begin{figure}[H]
\centering
\includegraphics[width=.9\textwidth]{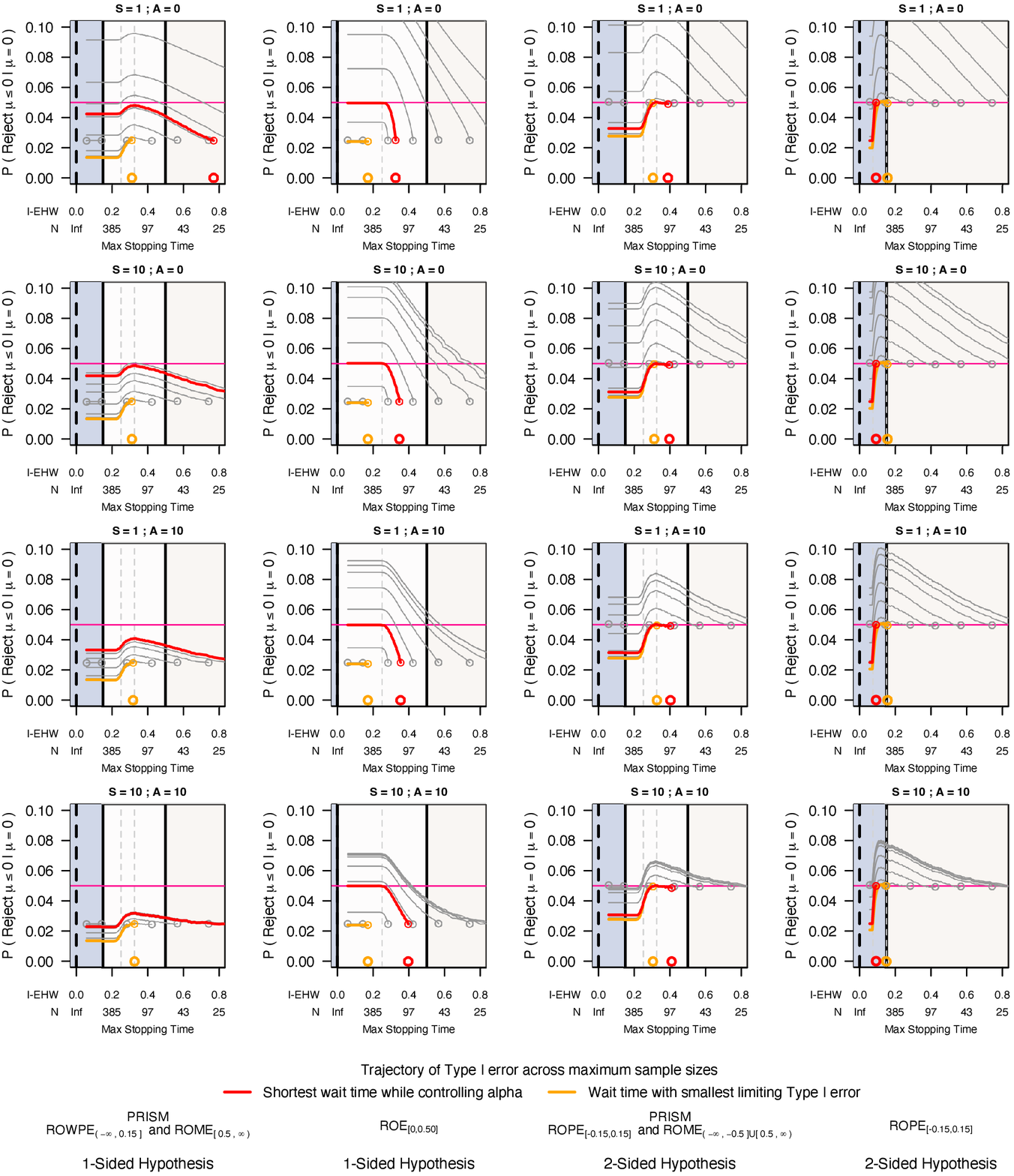}
\caption{\label{fig:supt1etrajectory}Type I error as a function of maximum sample size, monitoring steps, and affirmation steps. The x axis reads from right to left in terms of accruing data and left to right in terms of interval expected half-width (I-EHW). Circles denote the Type I error of a single assessment and Type I error increases for multiple assessments until a maximum sample size. Dashed lines reflect the midpoint between 0 and the positive ROME boundary and between ROPE and ROME positive boundaries.}
\end{figure}

Under these settings, monitoring the one-sided PRISM and ROE allowed for shorter minimum wait times than monitoring the two-sided PRISM and ROPE while controlling Type I error (Figure~\ref{fig:t1etrajectory}). While controlling Type I error, monitoring the one-sided PRISM had a smaller average sample size of 58 as compared to 158 when monitoring the one-sided ROE. Increasing $S$ and $A$ better controlled Type I error across all designs, though this generally did not decrease the wait time required to control Type I error. The exception was when monitoring the one-sided PRISM. In these simulations, Type I error was less than 0.035 when $S=10$ and $A=10$ regardless of the maximum sample size. Increasing to $A=10$ had a bigger impact upon controlling Type I error than increasing $S=10$ for all designs (Supplemental Figure~\ref{fig:supt1etrajectory}).

When monitoring PRISM and ROPE hypotheses, there was a point where Type I error decreased for increased maximum sample size. For PRISM monitoring, this point roughly occurred when accumulated data had an expected interval half-width of the Grey-Zone midpoint. The decrease in Type I error occurred because, at a maximum sample size, a set of trials were PRISM inconclusive yet make a Type I error. Yet, when followed until being SGPV conclusive some of these set of trials no longer ended as a Type I error. The same was true for ROPE designs and is true for non-null bound ROE designs. This suggests that PRISM monitoring may have added Type I error benefits under delayed outcomes.

\subsection{Monitoring frequency with delayed outcomes}
\label{subsec:delayed}

When outcomes are delayed, an adaptive monitoring design ideally arrives at a conclusion earlier (so as to reduce the required enrollment) and has minimal risk (ideally no risk) of contradicting itself when the delayed outcomes come to fruition. In a Type I error controlled study, two undesirable inconsistencies are to reject the null hypothesis then fail to reject the null hypothesis and vice-versa (succinctly described as "reversals" by \citep{hampson2013group}). While evidence on the PRISM helps guide overall conclusions under these inconsistencies, it is still worth investigating the properties of inconsistent null hypothesis conclusions under a PRISM design. In Section \ref{subsec:alphacontrol}, the monitoring the one-sided PRISM and ROE hypotheses had the greatest ability to control Type I error while being able to start monitoring earlier. Increasing $S$ and $A$ decreased Type I error, and we'd expect a similar benefit upon reversals when increasing $S$ and $A$. 

\begin{figure}[H]
\centering
\includegraphics[width=1\textwidth]{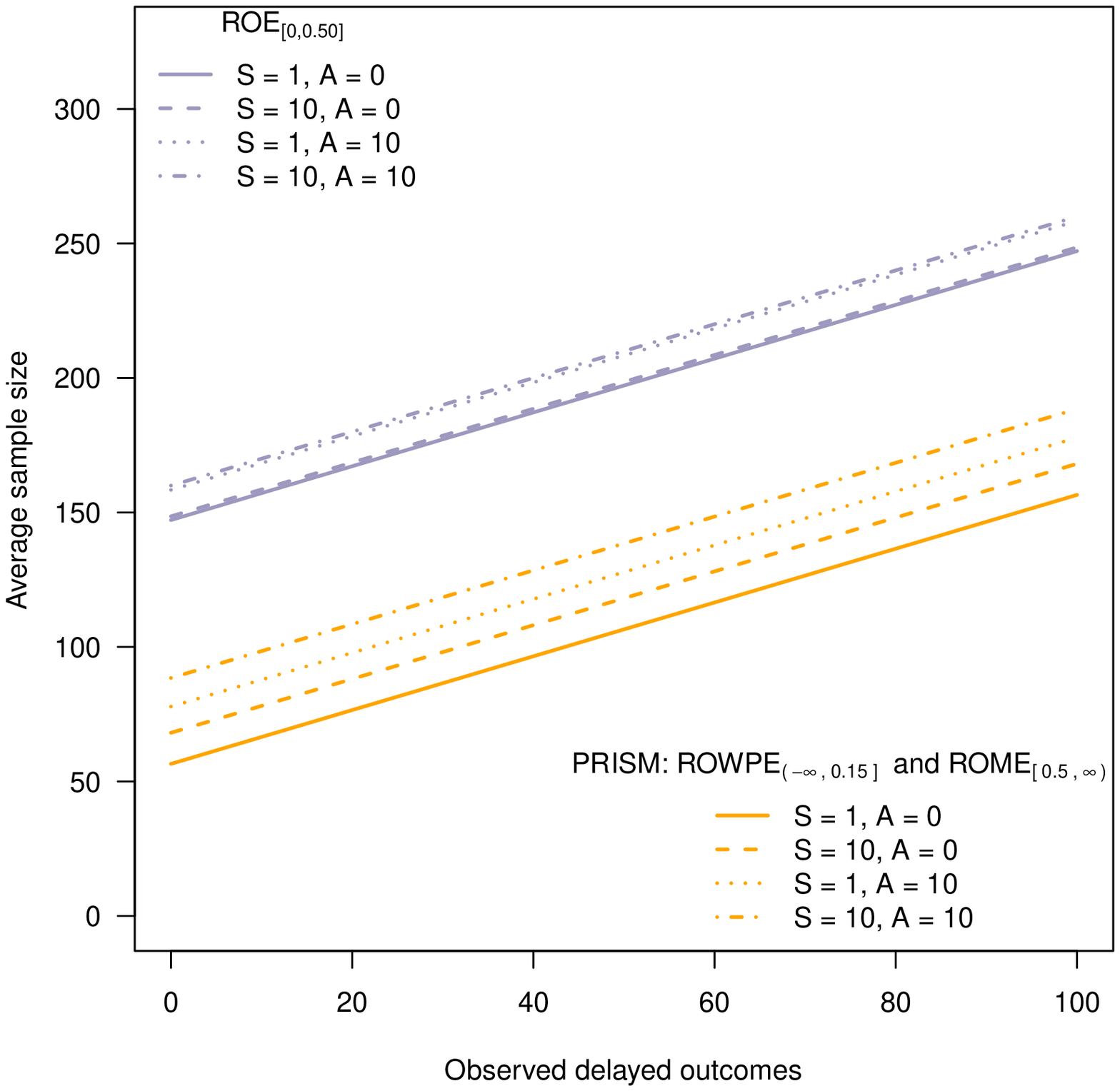}
\caption{\label{fig:suplagSampleSize}Average sample size under one-sided PRISM monitoring with $W = 20$ and $N = $ unrestricted and one-sided ROE monitoring with $W = 145$ and $N = $ unrestricted. These designs have limiting Type I error of 5\% under $S=1$, $A=0$, and instantaneous outcomes, and they are the designs used in studying delayed outcomes in Section~\ref{subsec:delayed}.}
\end{figure}

Following the aforementioned null-treatment effect simulation framework in Section~\ref{subsec:alphacontrol}, one-sided PRISM and ROE$_{[0,0.5]}$ hypotheses were monitored on 1 million mcmc replicates such that each design had a limiting Type I error of 0.05. The PRISM design was set with monitoring frequencies of $W$ = 20, $S$ = 1, $A$ = 0, and $N$ = unrestricted. The ROE design was set with $W$ = 145, $S$ = 1, $A$ = 0, and $N$ = unrestricted. After stopping due to monitoring rules, an additional 0 to 100 outcomes were observed as delayed outcomes. With $S=1$ and $A=0$, monitoring immediate outcomes using the one-sided PRISM required a third of the average sample size as monitoring the one-sided null-bound ROE (Supplemental Figure~\ref{fig:suplagSampleSize}). Keeping $W$ and $N$ the same, we increased $S$ and/or $A$ to 10 each. This decreases the limiting Type I error, and our interest was to see how it further controlled the risk of reversals. 

\begin{figure}[H]
\centering
\includegraphics[width=1\textwidth]{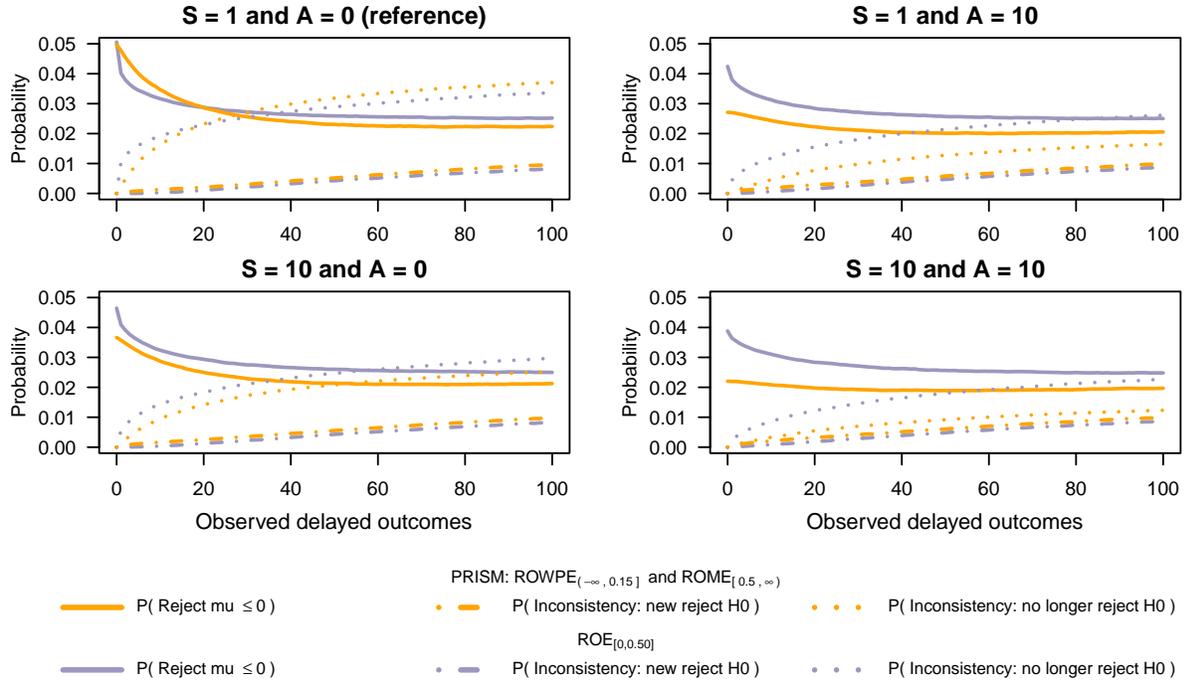}
\caption{\label{fig:lagOutcomes}Statistical significance reversal conclusions when monitoring one-sided PRISM and null-bound ROE and under null treatment effect.}
\end{figure}

In the reference setting ($S$=1, $A$=0), the ROE design advantageously reduced the overall Type I with few delayed outcomes (Figure~\ref{fig:lagOutcomes}). Yet, with increased delayed outcomes PRISM monitoring was more advantageous. This suggests that Type I errors under the PRISM design occurred when there was more evidence against the null hypothesis -- which is consistent with the PRISM's ROWPE. Increasing $S$ and $A$ reduced the risk of reversals in both designs though to a greater extent in PRISM monitoring. After 100 delayed outcomes, the total probability of a reversal decreased by a factor of 1.4 when monitoring PRISM versus the null-bound ROE. Increasing $A$ reduced the risk of a reversal more so than increasing $S$ under the null hypothesis, though at an increased average sample size (Supplemental Figure~\ref{subsec:delayed}).

\section{Real-world example}
\label{sec:reach}

\subsection{Context of REACH trial}

The Rapid Education/Encouragement And Communications for Health (REACH) randomized clinical trial \citep{Nelson:2018bw,mayberry2019text,mayberry2020mixed,Nelson2021} was designed to help adults with type 2 diabetes improve glycemic control (as measured by Percent Hemoglobin A1c -- HbA1c) and adhere to medication. Participants were randomized into one of three treatment arms with a 2:1:1 allocation ratio: enhanced treatment as usual (a text message when study HbA1c results are available), frequent diabetes self-care support text messages, and frequent diabetes self-care support text messages with monthly phone coaching. A meaningful, equally-allocated two-group comparison in this study was enhanced treatment as usual vs frequent self-care support. Follow-up assessment occurred at 3, 6, 12, and 15 months with 12 month HbA1c as the primary outcome. The study recruited about six participants a week and 300 a year.

The median REACH baseline HbA1c was 8.20 [IQR of 7.20, 9.53], and in this population, lower HbA1c reflects improved glycemic control. A change from baseline HbA1c of no more than +/- of 0.15 is practically equivalent to the point null of no change, whereas a decrease of HbA1c of 0.5 is meaningful to the point of adopting this novel intervention. The study was designed to test a two-sided hypothesis.

\subsection{Simulation}

We simulated the operating characteristics of adaptively monitoring the REACH trial for meaningful or non-meaningful effects (PRISM) using SeqSGPV. We simulated a trial assuming instantaneous outcomes using a two-sided PRISM ($H_0: \mu = 0$) and a one-sided PRISM ($H_0: \mu \ge 0$) for $N$ of 512, 650, and unrestricted. Then, we simulated a trial using a one-sided PRISM with 300 delayed outcomes for the same maximum sample sizes.

Using REACH data, we simulated designs which would control Type I error at 5\%. The two-sided PRISM was set with ROPE$_{[-0.15, 0.15]}$ and ROME$_{(-\infty, -0.5]\cup[0.5, \infty)}$ hypotheses with $W=420$, $S=25$, $A=0$, and $N=512$ to achieve a Type I error of 0.05. The one-sided PRISM was set with ROWPE$_{[-0.15, \infty)}$ and ROME$_{(-\infty, -0.5]}$ hypotheses with $W=55$, $S=25$, $A=0$, and $N=512$ to achieve a Type I error of 0.05 for instantaneous outcomes. To reduce Type I error and the risk of reversals, we also considered $W=55$, $S=25$, and $A=25$. An ordinary least square regression model was fit on accumulating data, and SeqSGPV monitored unadjusted 95\% confidence intervals estimating the treatment effect. 

For equally hypothesized treatment effects ranging from -1 to 1 by 0.025, we generated 120,000 bootstrap samples from REACH twelve-month HbA1c outcomes. Participants were assumed to have enrolled individually and were randomized in block two fashion to usual vs frequent self-care support. Outcomes followed the potential outcomes framework where, Y(0) equaled the 12 month outcome if randomized to the control arm and Y(1) equaled Y(0) plus the treatment effect if randomized to treatment. 

\subsection{Results}

\begin{figure}[H]
\centering
\includegraphics[width=1\textwidth]{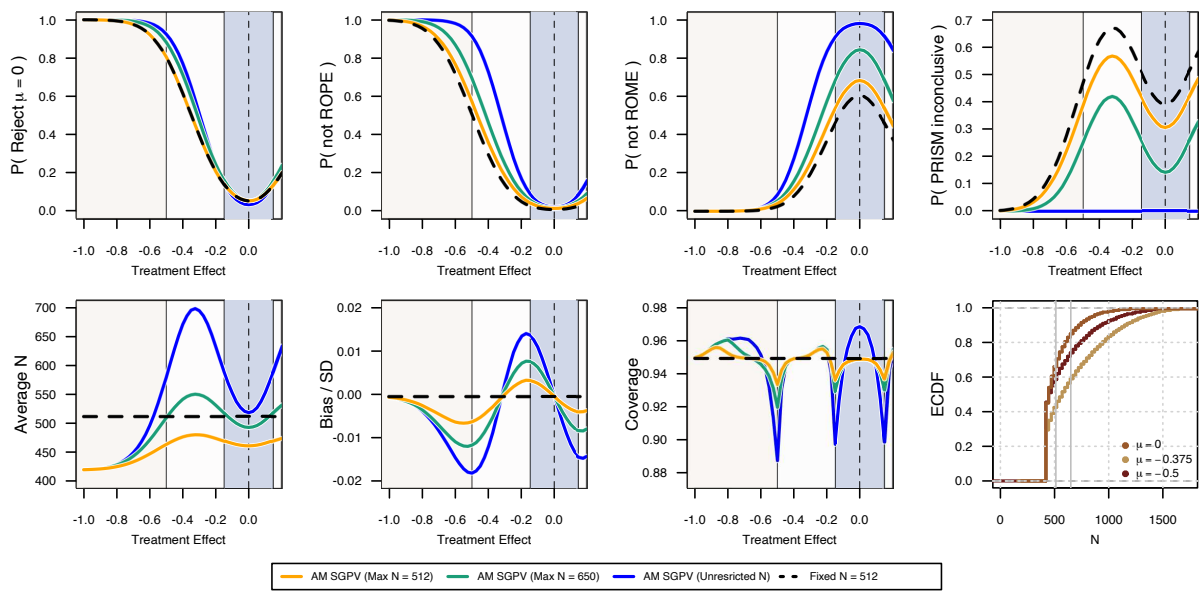}
\caption{\label{fig:oc2s}Operating characteristics from monitoring a two-sided PRISM on assumed immediate REACH trial outcomes.}
\end{figure}

Under the two-sided PRISM with immediate outcomes, the average sample size ranged from 421 to 480 under the constraint of $N = $512 (Figure~\ref{fig:oc2s}). This was sufficient for the study to be 81\% powered to detect a treatment effect of -0.5 and to have a Type I error of 0.05 -- which was equivalent to a single assessment at $N$ = 512. The chance of stopping earlier under treatment effects of 0 and -0.5 were, respectively 0.66 and 0.59. Average sample size was largest at the Grey Zone midpoint. Bias was no worse than an absolute standardized effect size of 0.02. There was no bias at the Grey Zone mid-point, and bias pulled toward the null for effects closer to the null and vice-versa. Compared to a single assessment at $N = 512$, the PRISM design was less likely to be PRISM inconclusive and more likely to find non-ROPE effects as non-ROPE and non-ROME effects as non-ROME.

\begin{figure}[H]
\centering
\includegraphics[width=1\textwidth]{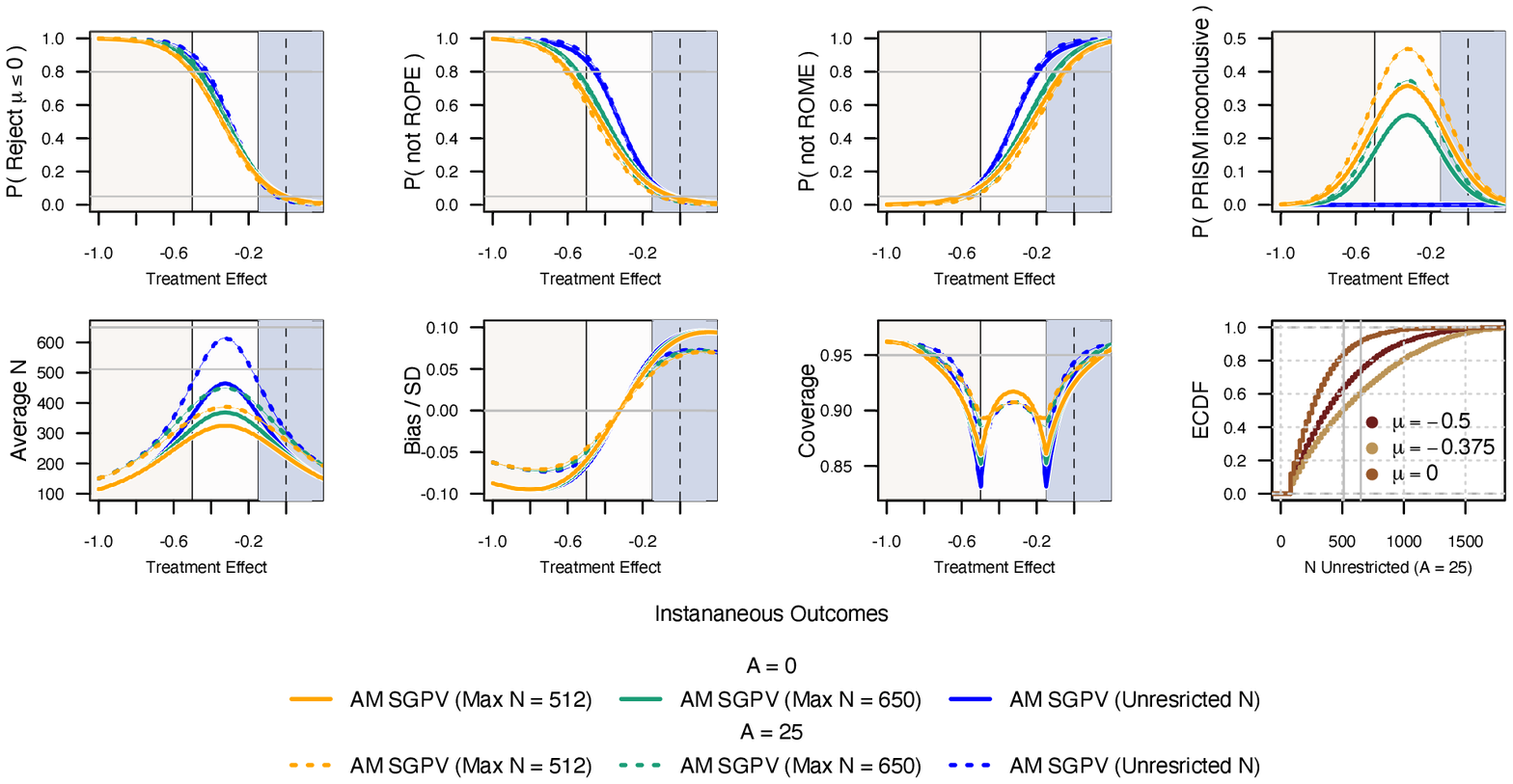}
\caption{\label{fig:REACH1s}Operating characteristics from monitoring a one-sided PRISM on data from the REACH trial assuming instantaneous outcomes.}
\end{figure}

Supposing immediate outcomes with a one-sided PRISM, the average sample size ranged from 115 to 325 under the constraint of $N=512$ and $A = 0$ (Supplemental Figure \ref{fig:REACH1s}). This was sufficient for the study to be similarly powered as the two-sided PRISM with equivalent Type I error. Undercoverage was most extreme at the ROWPE and ROME boundaries and bias was as much as 0.1 of an absolute standardized effect size. Requiring $A=25$ increased the average sample size yet reduced Type I error and bias. For unrestricted sample sizes, $A=25$ increased the probability finding non-ROPE effects as non-ROPE and non-ROME effects as non-ROME. However, under maximum sample sizes, $A=25$ increased the probability of ending PRISM inconclusive.

\begin{figure}[H]
\centering
\includegraphics[width=.9\textwidth]{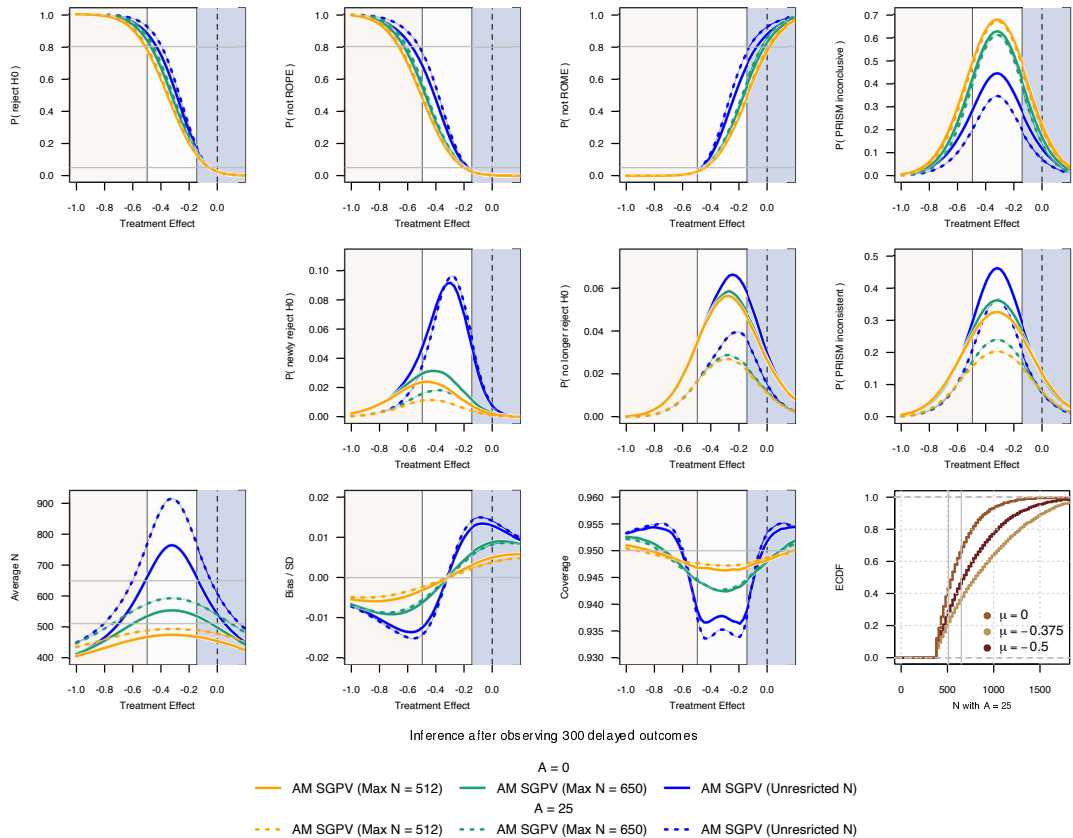}
\caption{\label{fig:oc1sLag300}Operating characteristics from monitoring a one-sided PRISM on data from the REACH trial in which outcomes were delayed by 300 observations.}
\end{figure}

In the real-life setting of 300 delayed outcomes, the average sample size for monitoring the one-sided PRISM ranged from 408 to 476 for $A=0$ and from 437 to 495 for $A=25$ (Figure~\ref{fig:oc1sLag300}). Due in part to fewer monitoring opportunities, the Type I error was 0.025 for $A=0$ and $A=25$ and power was 0.78 for a treatment effect of -0.5. The chance of stopping early under a treatment effect of 0 was 0.58 ($A=0$) and 0.44 ($A=25$). The affirmation step of $A=25$ decreased the risk of reversals from 0.03 ($A=0$) to 0.01 under the treatment effect of 0. For non-null effects, the chance of newly rejecting $H_0$ was less under $A=25$ than $A=0$ when under a maximum sample size of 512 and 650. With an unrestricted sample size, the chance of newly rejecting $H_0$ for a treatment effect at the Grey Zone mid-point was 0.08 for both $A=25$ and $A=0$.

\section{Conclusions and practical considerations}
\label{sec:conc}

We've proposed SeqSGPV as a novel, evidence-based sequential monitoring strategy using the Second Generation P-Value. SeqSGPV encompasses existing strategies to monitor indifferent zones using repeated credible and confidence intervals \citep{Hobbs:2008ce,Kruschke:2013jy,jennison1989interim}. Hence, our advancements include (i) an evidence-based and Type I error controlled monitoring scheme that uses the general purpose SGPV, (ii) a novel set of scientifically meaningful regions called the PRISM which encompass the null through a ROPE or ROWPE and includes a region of meaningful effects called ROME, and (iii) insight into using monitoring frequency to control Type I error, including a novel affirmation look to control Type I error.

The method-agnostic nature of the SGPV provides a universally available framework for measuring evidence on scientific hypotheses. The SGPV is quick and easy to calculate for any interval and composite hypothesis. For example, in these simulations, we monitored PRISM hypotheses on SGPV evidence using repeated 95\% confidence intervals. SeqSGPV makes multiple evidence assessments and a single assessment of rejecting the null hypothesis. Frequency properties are meaningful to asses for trial design \citep{Schonbrodt2018} though SeqSGPV does not guarantee exact interval coverage nor unbiased estimates across all treatment effects.

For designs achieving equivalent evidence and Type I error, our simulations showed benefit for monitoring PRISM over ROPE (two-sided hypothesis) and over null-bound ROEs (one-sided hypothesis). Monitoring PRISM hypotheses was dramatically more quickly conclusive than monitoring non-overlapping ROPE hypotheses. The same could be said of monitoring non-overlapping ROE hypotheses. PRISM's overlapping hypotheses close the expected sample size, which otherwise may result in indefinite monitoring for boundary effects. There may be a connection between the PRISM and the Triangular Test which we did not explore. When compared to a one-sided null-bound ROE with overlapping hypotheses, as done in our simulations, monitoring PRISM began more quickly, had a smaller average sample size, yielded better Type I error control, and reduced the risk of reversals under delayed outcomes.

For all designs considered, increasing $S$ and $A$ better controlled Type I error across possible maximum sample sizes. However, to achieve a nominal Type I error, increasing $S$ and $A$ had little, if any, impact upon wait time for monitoring the two-sided PRISM, ROPE, and null-bound ROE hypotheses.  There was a synergistic impact upon controlling Type I error through monitoring a one-sided PRISM and adjusting monitoring frequency/affirmations. Increasing $A$ controlled Type I error and reversals more strongly than increasing $S$, though with an increased average sample size. Further research could investigate trade-offs between increasing $S$ versus $A$.

In a real-world setting, we provided operating characteristics for monitoring a two-sided PRISM and one-sided PRISM. We considered the real-life delayed outcomes and imposed a maximum sample size of 512 (mimicking the actual trial’s maximum sample size and nature of delayed outcomes). After observing delayed outcomes, monitoring the one-sided PRISM achieved nearly the same power as a single assessment at $N=512$ and with half the Type I error. Under the null hypothesis, roughly half of simulated REACH trials would have stopped earlier than a total of $N=512$.

We recommend to first design a trial to have a desired Type I error under $S=1$ and $A=0$, then increase $S$ and $A$ as desired for a smaller overall Type I error. If feasible, we recommend monitoring a one-sided PRISM. In these simulations, monitoring the one-sided PRISM with a symmetric interval that devotes $\alpha/2$ to each tail increased the feasibility to design a trial with an early wait time while achieving a Type I error of size $\alpha$. In our simulations, a Type I error smaller than $\alpha$ was possible for any maximum sample size.

In practice, SeqSGPV requires simulation to design a study and assumed outcome variability. The R package, SeqSGPV, facilitates study design and includes additional, practical examples (\url{https://github.com/chipmanj/SeqSGPV}). As data accumulate, the sample variability should be continually assessed to ensure the study would not inflate the planned Type I error. Monitoring frequencies may be adapted as needed. Planning a trial to have less than the nominal Type I error may be a wise strategy in case outcome variability is greater than anticipated. Although we investigated the properties of SeqSGPV under delayed outcomes, one may consider using posterior predictive probabilities of success to determine whether to continue monitoring or not \citep{Saville:2014cg}. A possible, though undeveloped, method-agnostic strategy could entail a resampling strategy which supplements observed outcomes with resampled futuristic outcomes. When monitoring with confidence intervals, context of the possible bias and under-coverage should be provided in reporting results, or strategies should be taken to correct coverage (see for example \citep{jennison1989interim}). 

Broader conclusions regarding monitoring PRISM hypotheses and frequency are most directly applicable to evidence-based monitoring schemes such as the Sequential Probability Ratio Test, Sequential Bayes Factors, and repeated interval monitoring. Designs which tune or adjust stopping rules to achieve a maximum Type I error (such as tuning posterior probabilities required for efficacy/futility) are less directly comparable with the results in this work. Still, the use of PRISM monitoring and adjusting monitoring frequency may be applicable to such designs. As a concrete example, the BATTLE trial implicitly used a null-bound ROE with overlapping hypotheses to establish a treatment as non-null or non-meaningful. We showed a benefit of monitoring PRISM over a null-bound ROE but for designs having equivalent evidence. Future research could investigate the impact of PRISM versus null-bound ROE monitoring when using tuned or adjusted evidence levels for stopping rules.

\bibliographystyle{agsm}
\bibliography{2022_04_refs}

\pagebreak
\bigskip
\begin{center}
{\large\bf SUPPLEMENTARY MATERIAL}
\end{center}

\beginsupplement

\begin{figure}[H]
\centering
\includegraphics{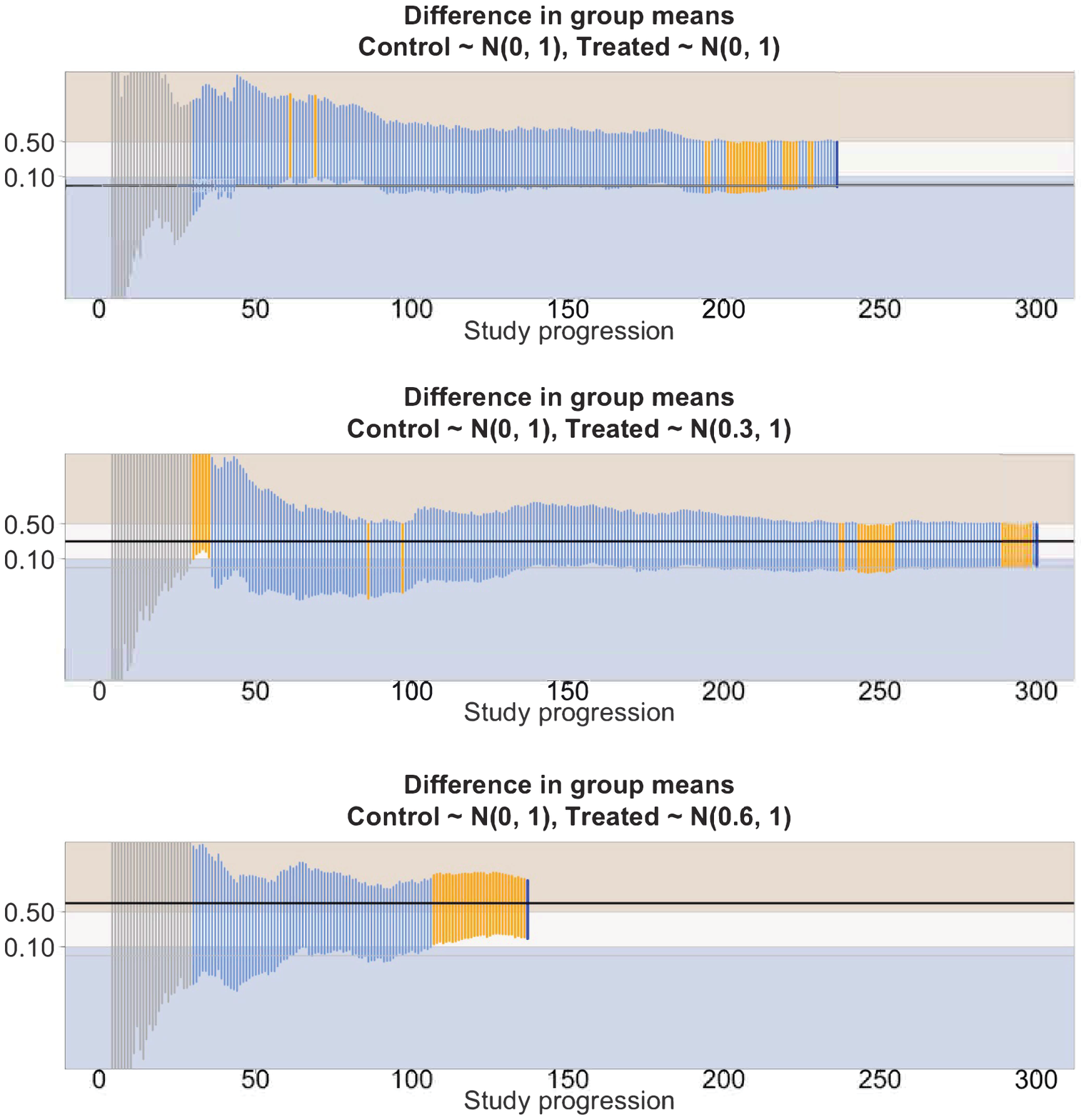}
\caption{\label{fig:supExamples}Examples of SeqSGPV monitoring of one-sided PRISM hypotheses with $W=30$, $S=1$, $A=30$, and $N=$ unrestricted. Grey intervals ($W<30$) are not evaluated for evidence. Orange intervals raise a stopping indication alert. Final dark blue interval occurs when affirming an alert.}
\end{figure}

\end{document}